\documentclass[sigconf,natbib=false,nonacm]{acmart}

\AtBeginDocument{%
  }

\RequirePackage[
  datamodel=acmdatamodel,
  style=acmnumeric,backend=biber
  ]{biblatex}

\usepackage[inline]{enumitem}
\usepackage{booktabs}
\usepackage{subcaption}
\usepackage{balance}

\renewcommand{\texttt}[1]{%
  \begingroup
  \ttfamily
  \begingroup\lccode`~=`/\lowercase{\endgroup\def~}{/\discretionary{}{}{}}%
  \begingroup\lccode`~=`.\lowercase{\endgroup\def~}{.\discretionary{}{}{}}%
  \catcode`/=\active\catcode`[=\active\catcode`.=\active
  \scantokens{#1\noexpand}%
  \endgroup
}

\begin{document}

\title{OpenDORS: A dataset of openly referenced open research software}

\author{Stephan Druskat}
\affiliation{%
  \institution{German Aerospace Center (DLR)}
  \city{Berlin}
  \country{Germany}
}
\email{stephan.druskat@dlr.de}
\orcid{0000-0003-4925-7248}

\author{Lars Grunske}
\affiliation{%
  \institution{Humboldt Universit\"at zu Berlin}
  \city{Berlin}
  \country{Germany}
}
\email{grunske@informatik.hu-berlin.de}
\orcid{0000-0002-8747-3745}


\begin{abstract}
  In many academic disciplines, software is created during the research process 
  or for a research purpose. 
  The crucial role of software for research is increasingly acknowledged. 
  The application of software engineering to research software has been formalized as 
  research software engineering, to create better software that enables better research. 
  Despite this, large-scale studies of research software and its development are still lacking. 
  To enable such studies, we present a dataset 
  of 134,352 unique open research software projects and 134,154 source code repositories 
  referenced in open access literature. Each dataset record identifies 
  the referencing publication and lists source code repositories of the software project. 
  For 122,425 source code repositories, the dataset provides metadata on latest versions, 
  license information, programming languages and descriptive metadata files. 
  We summarize the distributions of these features in the dataset and describe additional 
  software metadata that extends the dataset in future work. Finally, we suggest examples 
  of research that could use the dataset to develop a better understanding of 
  research software practice in RSE research.
\end{abstract}




\maketitle

\section{Introduction}

Software has become a central building block in research across many disciplines.
``Research software''~\cite{gruenpeter_defining_2021} is created as part of the research process or specifically for the purpose of enabling and improving research.
The quality of research software was regarded as negligible compared to the quality of software applied in military or business contexts~\cite{randell_fifty_2018}.
Today, research software is increasingly acknowledged as being critical for developing new technologies, expanding knowledge and addressing grand challenges.
The field of research software engineering is continuously formalized through the establishment of the research software engineer (RSE) role, institutional RSE units and national associations.
There is growing interest in research software within the software engineering research community,
and in opportunities for RSE research~\cite{felderer_investigating_2025}
that investigates 
features, challenges and methods, and contextual requirements specific to research software and its engineering.
RSE research can help understand research software as a distinct class, and develop innovative solutions and integrations of software engineering processes, methods and tools in research software development.
This will ultimately lead to the creation and use of better software for more and better critical research.
Consequently, premier computer science and software engineering venues host events that focus on research software engineering, e.g., Dagstuhl Seminar ``Research Software Engineering: Bridging Knowledge Gaps''\cite{druskat_research_2025} (2024) and the workshop ``Software Engineering and Research Software''\footnote{\url{https://conf.researchr.org/home/icse-2026/sers-2026}} (SERS '26) at ICSE.
Work by the \textit{RSE community} comprises technical papers, position papers~\cite{anzt_environment_2021,jay_software_2021} and practice papers discussing research software engineering~\cite{druskat_better_2025,druskat_software_2024}.
One focus of \textit{software engineering research on research software} is qualitative and smaller scale quantitative work to understand aims and modes of research software production~\cite{segal_software_2009,kelly_scientific_2015}, or unique challenges for engineering domain-specific research software~\cite{vogel_challenges_2019}.
Quantitative methods such as larger surveys ($N>1,000$) are used to describe the state of the practice, e.g., for testing~\cite{eisty_testing_2025}, regional communities~\cite{carver_survey_2022} or professional backgrounds and attitudes towards software engineering concepts~\cite{hannay_how_2009}.
Other work mixed mining software repositories with surveys (Milewicz et al.~\cite{milewicz_characterizing_2019} for contribution roles), or with qualitative analysis (Pimentel et al.~\cite{pimentel_understanding_2021} for computational notebooks).
Larger and large scale work mined software repositories for computational notebooks~\cite{pimentel_large-scale_2019,wang_assessing_2021} or investigate software metadata~\cite{hounsri_good_2025}.
There is a lack of large scale and more comprehensive empirical studies of research software specifics and of its commonalities with other software classes across categories~\cite{hasselbring_multidimensional_2025}, programming languages, project sizes, research domains, etc.
This is due to a lack of large, sufficiently representative research software datasets that provide access to repositories for mining.
To fill this latter gap, we present \textit{OpenDORS: Open Dataset of Openly Referenced Open Research Software}, a dataset of 134,352 unique research software projects and 134,154 unique source code repositories.
We release the dataset and the workflow to construct it under an open license~\cite{druskat_opendors}.
We also release a Python package \texttt{opendors}\footnote{\url{https://pypi.org/project/opendors/}} that implements the construction logic as open source code~\cite{druskat_opendorslib}.
The workflow and package, together with openly licensed publication metadata,
make the dataset reproducible.

\section{Related work}

Related work discusses research software identification and can be grouped into the areas citation, mentioning and strategies for identification through document linking or LLM use.
Formal \textit{software citation}~\cite{smith_software_2016} makes it possible to identify research software through generic citation analysis.
However, current academic practice often falls short of the principles of software citation.
Instead, used software is often informally mentioned, missing relevant metadata such as URLs to artifacts.
This complicates automated retrieval of software artifact URLs~\cite{howison_software_2016,sandt_practice_2019,du_understanding_2022,druskat_dont_2024}.
In lieu of good software citation practice, recent related work has focused on analyzing \textit{mentions to software} in the literature.
SoftCite~\cite{du_softcite_2021} provides metadata for 4,093 software mentions in biomedicine and economics literature, manually annotated by 36 students over 2 years.
It includes 2,541 sets of metadata that include URLs to websites related to the software.
While this approach may yield high quality metadata, it does not scale to larger scale datasets.
The SoftCite dataset was used to train a machine-learning model that was used to build the CORD-19 Software Mentions~\cite{wade_cord-19_2021}.
It contains software names mentioned in over 77,000 COVID-19-related full-text articles,
but no URLs to access software artifacts.
SoMeSci~\cite{schindler_somesci_2021} consists of 3,756 manually annotated software mentions from 1,367 articles, around 40\% of which provide a URL
of~``a location where additional information can be obtained''~\cite[p. 4577]{schindler_somesci-_2021-1}.
Finally, CZ Software Mentions~\cite{istrate_cz_2022} includes $\sim19.3M$ software mentions (2,536,801 unique mentions) from $\sim20.7M$ publications.
Mentions were mapped to $\sim185,000$ URLs for five software ecosystems (PyPI, CRAN, Bioconductor, SciCrunch, GitHub) using string matching,
which greatly reduces linking accuracy~\cite{brown_biomedical_2025,druskat_dont_2024}.
This makes the dataset unfit for
large-scale empirical studies of research software through mining software repositories.
Datasets that \textit{link research software} mentioned in the literature
to related software artifacts can be constructed using URL extraction.
This is a promising approach, as issues with linking accuracy are largely irrelevant.
Lin et al.~\cite{lin_automatic_2022} provide a dataset of 7,502 URLs detected in regular papers presented at 10 AI conferences between 2010 and 2019
using machine-learning classifiers.
However, over a quarter of URLs are not to source code repositories or other software artifacts.
Garijo et al.~\cite{garijo_bidirectional_2024} provide a dataset of 1,485 research software linked to their implementations.
Their data is based on software mentions extracted from over 14,000 software engineering preprints on the ArXiv
for which reciprocal links exist between preprint and respective source code repository.
As this dataset requires bidirectional mentions to exist, the approach may fail to identify unidirectional mentions,
and will not contain data for software that is related work rather than implementation of the research that the preprint reports.
RepoFromPaper~\cite{stankovski_repofrompaper_2024} aims to alleviate this issue through the use of sentence classification models
on 1,800 artificial intelligence preprints on the ArXiv.
URLs are extracted from sentences that rank high in machine-learning classification of whether they contain implementation links.
While this approach does not require reciprocal links between preprint and source code repository,
the scope of this research remains limited to implementation code for published research.
The resulting dataset contains 604 URLs of implementation code.
SciCat~\cite{malviya-thakur_scicat_2023} discounts backlinks from repositories to publications and 
investigates an alternative method to mention linking and URL extraction.
An \textit{LLM} is used to identify scientific application software through prompting questions against 354,943 \texttt{README.md} files found in a subset of 430,469 repositories filtered from World of Code~\cite{ma_world_2021}.
The method validation shows significant shortcomings of this method in the identification of research software types.
A human-annotated sample of 60 repositories yields strong recall, but very low agreement between the LLM and the human annotators (Cohen's $\kappa=0.150$).
Cohen's $\kappa=0.130$ for the agreement between human annotators points to inherent difficulty in identifying scientific application software.
Additionally, precision is low ($0.4$), yielding 60\% false positives on the sample.
So far, identifying software artifact URLs based on software mentions has not provided reliable results.
The results of utilizing large language models to identify research software in known source code repositories suggest
that more research is needed to create large datasets of research software using this method.
We have used a different approach to construct the OpenDORS dataset.
We utilize URL extraction, but circumvent the detour via software mentions.
Instead, we use component analysis of parsed URLs to filter source code repository URLs extracted from publications.

\section{Dataset construction}

\begin{figure}[h]
  \centering
  \includegraphics[width=\columnwidth]{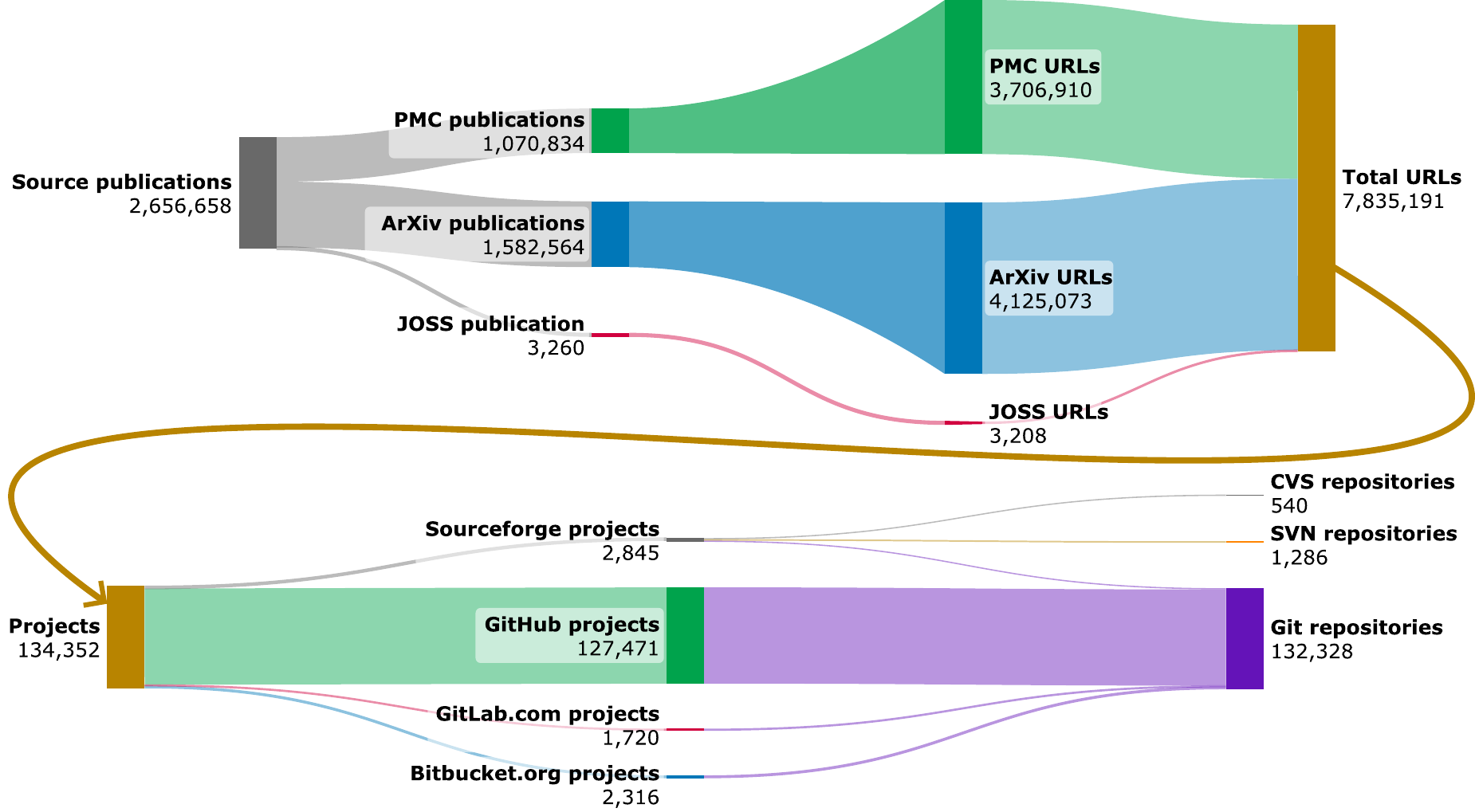}
  \caption{\small Data provenance over the dataset construction process.}
  \Description{Sankey diagram showing the outcomes of the single steps of the dataset construction workflow.}
  \label{fig:process}
\end{figure}

Data provenance for OpenDORS is visualized in Figure~\ref{fig:process}.
The single construction steps are detailed below. The process is designed according to the following criteria:
\begin{enumerate*}[label=(\alph*)]
    \item \textbf{attestation:} repositories referenced in openly accessible publications are included independently of their availability and accessibility, mentioning publications are identifiable;
    \item \textbf{historicity:} mentioning publications are linked to likely referenced software versions;
    \item \textbf{diversity:} repositories using different version control systems (Git, Subversion) on different coding platforms (\texttt{github.com}, \texttt{gitlab.com}, \texttt{bitbucket.org}, \texttt{sourceforge.net}) are included;
    \item \textbf{realistic code sharing:} repositories are included independently of whether they are formally and correctly licensed;
    \item \textbf{machine-actionability:} where available, machine-actionable identifiers are provided that allow direct access to source code;
    \item \textbf{discoverability:} minimally viable metadata is provided for repositories to assess relevance for sampling.
\end{enumerate*}
Based on these criteria, we include different sources for publication data and metadata.
The process includes the following high-level steps.

\textit{1) URL collection and processing, metadata enrichment, record creation:}
We build on a dataset of URLs extracted from publication PDFs from PMC (1900-2022) and ArXiv (04/2007-12/2021)~\cite{escamilla_extract-urls} to retrieve records of 3,706,910 raw URLs parsed from 1,070,834 PMC publications and 4,125,073 raw URLs parsed from 1,582,564 ArXiv publications.
We also retrieve 3,208 URLs from 3,260 publications in JOSS (up to and including 2025-11-05) through its JSON interface.
We analyze URL components to find URLs for source code repository platforms, to canonicalize URLs and to infer potential version information.
Given a URL schema for a source code repository platform and the components
of a mentioned URL, we can determine the platform through the network location component.
We can also identify potential version information implied by the URL path.
URLs for \texttt{bitbucket.org} repositories, for example, define the location of Git branches with \texttt{?at=} queries.
Similarly, all exclusively Git-based platforms define the location of tags by preceding tag identifiers
with of \texttt{tag}/\texttt{tags} and/or \texttt{release}/\texttt{releases} path components.
URLs locating branches in Subversion repositories on SourceForge always contain the path components \texttt{tree/branches}, etc.
We enrich mention metadata with publication dates retrieved from two pre-prepared metadata sets providing publication dates for PMC and ArXiv publications~\cite{druskat_publication_2024,druskat_arxiv_2024}.
We also retrieve archive information -- including specific software version identifiers --
from JOSS metadata.
For each canonical URL, we create OpenDORS research software records, including mention metadata and platform.

\textit{2) Mining SourceForge repository data:}
SourceForge projects can contain zero or more Git, Subversion, Mercurial or CVS source code repositories.
We merge all records for research software projects on \texttt{sourceforge.net} and
retrieve information about available repositories for each from the SourceForge Allura API, including URLs for anonymous cloning.
For all other platforms, we construct clone URLs based on the platforms' URL schemas.

\textit{3) Source code repository cloning and processing:}
We create separate files for each research software record,
and attempt to clone each recorded Git and Subversion repository.
We record if the repository is accessible.
GitLab projects can contain zero or more subprojects and one or more Git repositories.
For all projects where cloning with the previously  constructed clone URL fails,
we determine the structure of projects on \texttt{gitlab.com} via the REST API,
restructure the research software record, and attempt to determine a main
repository for the mentioned project through string matching heuristics.
This differs from the treatment of SourceForge projects, which lack the semantically relevant notion of subprojects.
Instead, each repository on \texttt{sourceforge.net} is treated as part of the same project,
because
\begin{enumerate*}[label=(\alph*)]
    \item repositories may contain different versions of the same codebase, e.g., a CVS repository that was migrated to a Subversion repository that was in turn migrated to a Git repository, as version control systems are modernized;
    \item the structure of Subversion repositories is recommended but not enforced and the repositories for separate software projects, or for modules of the same project may live in the same repository root, where it is hard to discern them from each other;
    \item the structural properties of Subversion repositories can lead to erratic nesting
    of source code roots, which we found to be the case through manual analysis of a subset of repositories for which adherence to the recommended repository structure with \texttt{trunk}, \texttt{branches} and \texttt{tags} directories could not be determined programmatically.
\end{enumerate*}
Despite this, we attempt to identify the repository that is most likely the mentioned one through string matching heuristics.

We analyze the structure and history of each successfully cloned repository to identify 
\begin{enumerate*}[label=(\roman*)]
    \item the latest version of the software contained in the source code repository at the time of mining the source code repository;
    \item the latest version preceding the publication date of each publication mentioning the research software, where no such version information already exists, e.g., from
    archive metadata for JOSS publications.
\end{enumerate*}
For each version to determine, we record the highest ranking available version
that we determine based on version type and information inferred from mentioning URLs.
Version types are ordered by their intentionality and by the visibility of the version.
Tags indicate intended proactive versioning and rank highest.
Specified revisions or commits are well-defined versioning information.
The inclusion of revision or commit identifiers in a URL mentioning a research software in a publication implies intention.
If URLs include branch information, this is less specific than revision/commit information
and less highly ranked, but still indicates a degree of intentionality.
For this type of version, the latest tag preceding the publication date of the mentioning publication that is included in the branch history is recorded,
or in lieu of tags, the latest commit.
The version information of URLs pointing to specific paths is less reliable than branch
information, but indicates a residue of intentionality.
For path-based versions, the latest tag or revision/commit is recorded in the same way as for branches.
The lowest ranking version type includes named versions, e.g., 
as provided in the metadata of an archived version of the repository.
If the version name does not correspond to a tag or revision identifier,
the version information is considered non-deterministic and name-based versions thus rank lowest.
For each successfully identified latest version of a repository at the time of mining it,
we detect license information (using \texttt{licensee}~\cite{licensee}), 
distribution of programming language artifacts (using \texttt{linguist}~\cite{linguist})
and files providing generic or citation-specific software metadata
(\texttt{CITATION.cff}, \texttt{codemeta.json}, \texttt{.zenodo.json}).
Finally, we merge all research software records to produce the dataset.

\begin{table}
  \caption{\small Quantitative description of \textit{OpenDORS v2025-11}.}
  \label{tab:projects}
  \resizebox{\columnwidth}{!}{
      \begin{tabular}{rccccc}
        \toprule
        \textbf{Research software projects} & $\geq1$ repository & No repository & \textbf{Total} & & \\
         & 133,193 & 1,159 & \textbf{134,352} & & \\
        \midrule
        \textbf{Projects per platform} & \texttt{github.com} & \texttt{gitlab.com} & \texttt{bitbucket.org} & \texttt{sourceforge.net} & \\
         & 127,471 & 1,720 & 2,316 & 2,845 & \\
        \midrule
        \textbf{Source code repositories} & Git & Subversion & CVS & \textbf{Total} & Avg./project \\
         & 132,328 & 1,286 & 540 & \textbf{134,154} & 1.997 \\
        \cmidrule{2-6}
        \textbf{Latest repository version} & Retrieved & Not retrieved & & &  \\
         & \textbf{122,425} & 11,729 & & &  \\
        \midrule
        \textbf{Software mentions} & PubMed Central & ArXiv & JOSS & \textbf{Total} & Avg./project \\
         & 19,716 & 213,444 & 3,208 & \textbf{236,368} & 1.759 \\
        \midrule
        \textbf{Version types} & \texttt{tag} & \texttt{revision} & \texttt{branch} & \texttt{path} & \texttt{name} \\
         & 73,784 & 120,993 & 12,574 & 569 & 27 \\
        \midrule
        \textbf{License types} & Permissive & Strong copyleft & Weak copyleft & Other & None \\
         (avg. confidence 99.281) & 46,345 & 15,027 & 2,217 & 467 & 160 \\
        \midrule
        \textbf{Metadata files} & \texttt{CITATION.cff} & \texttt{.zenodo.json} & \texttt{codemeta.json} &  &  \\
         & 1,975 & 468 & 297 &  &  \\
        \bottomrule
      \end{tabular}
  }
\end{table}

\textit{Implementation:}
Our dataset construction approach combines 
different types of data sources (source code repositories, existing datasets, web resources and APIs), 
different methods and
tools from different programming language and packaging ecosystems
(version control packages, Python wrappers, data analysis tools in Ruby gems).
It also requires our own implementation of data processing and heuristics,
which we provide in an open source Python package, \texttt{opendors}~\cite{druskat_opendorslib}.
While file-based data processing tasks
can be run efficiently in parallel,
some tasks require access to resources through rate-limited web APIs.
We use and release a Snakemake~\cite{snakemake_9_13_5,molder_etal_2025_sustainable_data_analysis} workflow~\cite{druskat_opendors} 
and different \texttt{conda}
environment definitions 
to orchestrate this diversity of data, tools and methods
and support the reproducibility of the dataset.
For the current version of OpenDORS, we run the Snakemake workflow on a small high-performance data analysis cluster
at the German Aerospace Center's Institute for Software Technology.

\section{The OpenDORS dataset}

\begin{figure}[h]
  \centering
  \includegraphics[width=\columnwidth]{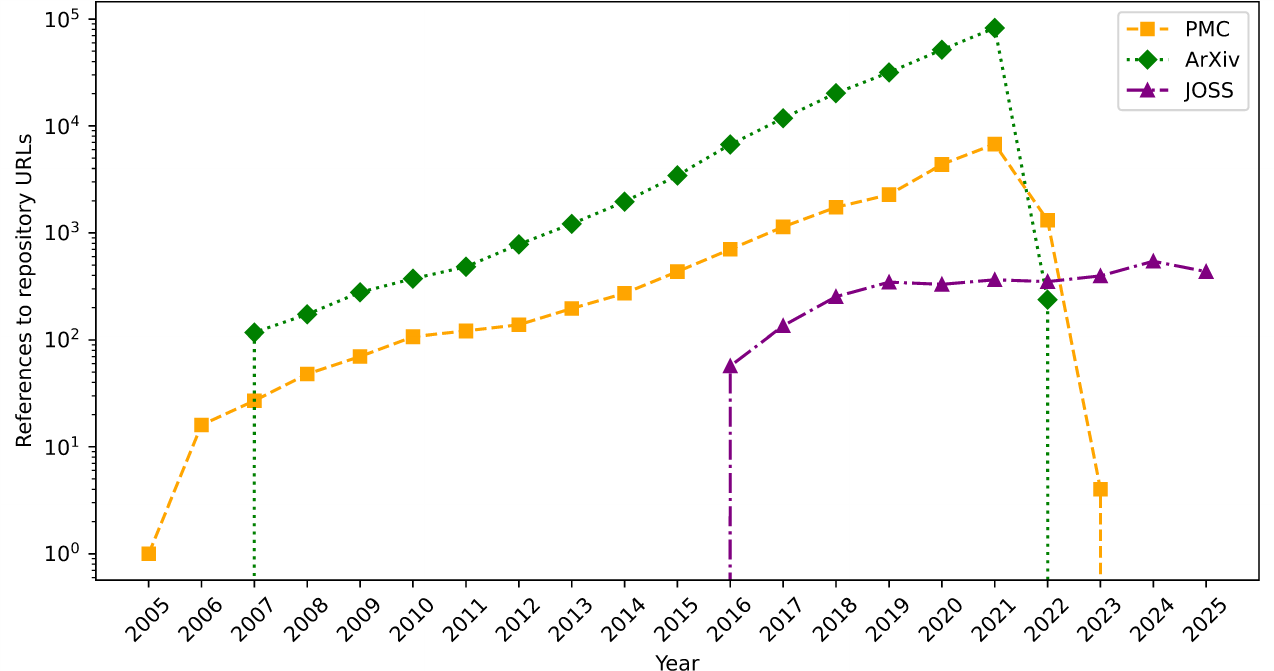}
  \caption{\small References (log scale) to repositories per data source. Declines in PMC/ArXiv references reflect publications available in~\cite{escamilla_extract-urls}.}
  \Description{A line graph showing the number of references to source code repository URLs per year for three different data sources (PMC, ArXiv, JOSS).}
  \label{fig:urls}
\end{figure}

\begin{figure}[h]
  \centering
  \includegraphics[width=\columnwidth]{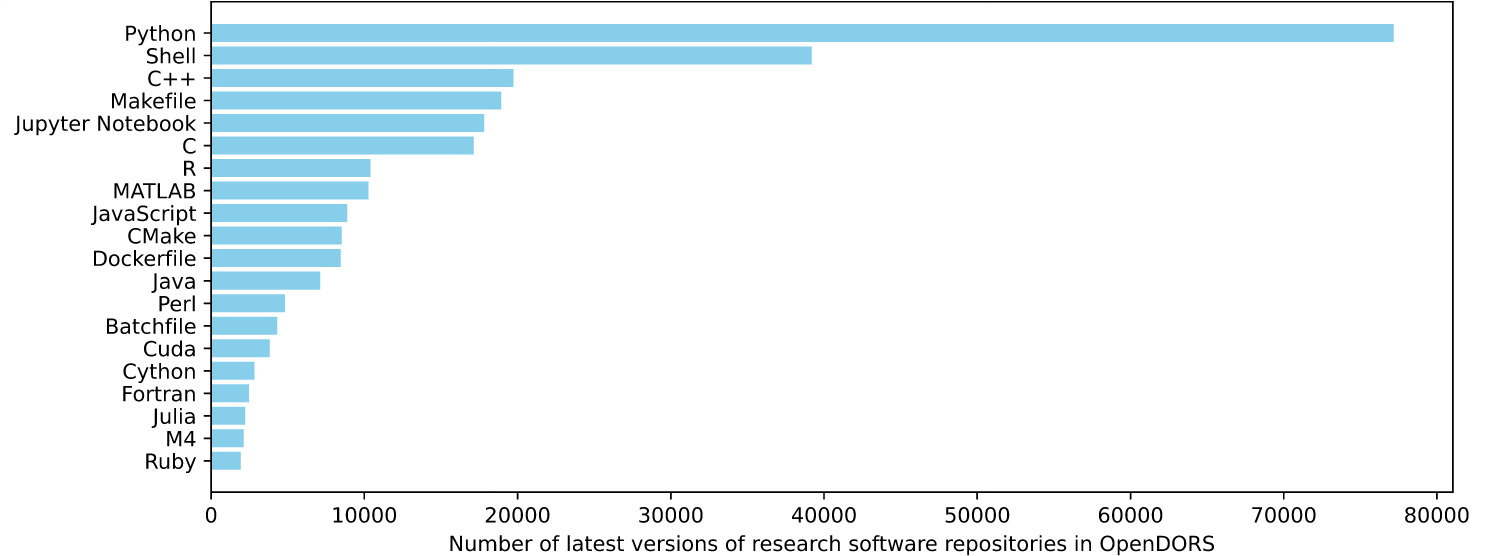}
  \caption{\small Repository counts for programming languages (top 20).}
  \Description{A bar chart showing for 20 programming languages how many source code repositories contain code in that language.}
  \label{fig:languages}
\end{figure}

\textit{OpenDORS v2025-11} contains a collection of
134,352 research software projects
based on repository URLs mined from 2,656,582 publications in 
PubMed Central\footnote{\url{https://pmc.ncbi.nlm.nih.gov/}} (PMC), 
ArXiv\footnote{\url{https://arxiv.org/}} and 
the Journal of Open Source Software~\footnote{\url{https://joss.theoj.org}} (JOSS).
It includes metadata for mentioning publications, for related source code repositories and for latest software versions.
In total, OpenDORS contains 134,154 unique source code repositories with 122,425 latest versions.
We provide further descriptive details in Table~\ref{tab:projects}.
Figure~\ref{fig:urls} shows the distribution of referenced repository URLs in our data sources over time.
Figure~\ref{fig:languages} gives the number of repositories containing code in specific programming languages.
The 144MB dataset JSON file \texttt{OpenDORS.v2025-11.json} is provided in a 16MB xz-compressed tarball \texttt{OpenDORS.v2025-11.tar.xz}.
The JSON-serialized model schema for the OpenDORS dataset is provided in the \texttt{schema.json} file.

\section{Research opportunities}

The OpenDORS dataset provides a unique resource for conducting large-scale empirical software engineering research into research software practices, i.e., RSE research~\cite{felderer_investigating_2025},
through mining software repositories.
It can generally be used to create evidence towards validating or refuting claims in the literature with respect to research software and its engineering that are currently lacking strong empirical foundations. Combined with existing datasets for other classes of software, OpenDORS can also be used for contrastive studies of different software engineering communities and their contexts, such as makeup and changeover rates of development teams, professional backgrounds and experience, software quality and lifecycles, and uniqueness of domain problems and related challenges.
Focusing on the research context, OpenDORS provides opportunity for large scale reproducibility studies, classification tasks identifying research purpose, uptake of software engineering techniques, derivation of required skills for RSEs, identification of research software lifecycle patterns, and many others.
On a meta-level, the dataset can help gain insights into the impact of funding or institutional policies on research software and software quality and, vice versa, the influence of research software on the impact of research institutions, communities or individual RSEs and researchers.

\section{Limitations}

Some factors limit the usefulness of the current version of the OpenDORS dataset for specific research questions.
Our data source cover a limited subset of research domains.
This hinders RSE research with a domain-specific focus
beyond the biomedical and life sciences, physics, mathematics and computer science.
The scope of the dataset is restricted to openly available software.
Therefore, it cannot be used to investigate research software
that is closed source, or remains unshared.
Here, it misses to provide opportunities for comparative empirical research taken different sharing modes and requirements into account.
On a more technical level, not all referenced repositories recorded in the dataset are guaranteed to contain software.
Some may be used to develop websites or other output.
The semantics of references to research software in our dataset are not entirely clear, as citation intent is not analyzed during the mining process.
Finally, our focus on URLs for large public repository platforms excludes research software that is being shared somewhere else, for example on platform instance run by research institutions.

\section{Conclusion and future work}

\textit{OpenDORS v2025-11} is the largest actionable dataset to date that reliably links references to research software to source code repositories.
It represents a promising resource for empirical RSE research at scale.
To address known limitations of the dataset and its construction process,
future work includes extending and updating the collection of data sources.
Improved filtering should be used to exclude non-software repositories.
Research domain metadata for publications should be mined to allow domain-specific insights.
Versioning metadata should be extended to allow for better understanding of research software lifecycles.
Finally, identified latest versions should be annotated with identifiers for their archived versions in the Software Heritage Archive.
\printbibliography

\end{document}